\documentclass[12pt]{JHEP3}

\usepackage{amssymb,amsmath}
\usepackage{graphics}
\usepackage{epsfig}
\usepackage{amsfonts}

\usepackage{psfrag}

\usepackage{amscd}

\newcommand{\be}[1]{ \begin{equation}\label{#1} }
\newcommand{\ee}{\end{equation}}
\newcommand{\bea}[1]{\begin{eqnarray}\label{#1} }
\newcommand{\eea}{\end{eqnarray}}
\newcommand{\eq}[1]{(\ref{#1})}

\newcommand{\idn}{{1\relax{\kern-.35em}1}}

\newcommand{\sn}{{\rm sn}}
\newcommand{\cn}{{\rm cn}}
\newcommand{\dn}{{\rm dn}}

\title{Worldsheet Properties of Extremal Correlators in AdS/CFT}
\author{Justin R. David\footnote{On lien from Harish-Chandra Research Institute, Allahabad.}
$^{a}$, Rajesh Gopakumar$^b$, Ayan Mukhopadhyay$^b$\\
$^a$Centre for High Energy Physics, \\ ~Indian Institute of Science, \\~Bangalore 560012, India. \\
$^b$Harish-Chandra Research Institute, \\ ~Chhatnag Road, Jhunsi,\\
~Allahabad 211019, India. \\
\email{justin@cts.iisc.ernet.in}, \email{gopakumr@hri.res.in},
\email{ayan@hri.res.in}}

\abstract{We continue to investigate planar four point worldsheet correlators of string theories which
are conjectured to be duals of free gauge theories. We focus on the extremal correlators    $\langle {\rm { Tr}}(Z^{J_1}(x))
{\rm{Tr}} (Z^{J_2}(y)){ \rm{ Tr}}(Z^{J_3}(z)){\rm{ Tr}}(\bar{Z}^{J}(0)))\rangle$ of $\mathcal{N} = 4$ SYM theory, and construct
the corresponding worldsheet correlators in the limit when the $J_i \gg 1$. 
The worldsheet correlator gets contributions, in this limit,  from a whole family of Feynman graphs. 
We find that it is supported on a {\it curve} in the moduli space parametrised by the worldsheet crossratio. In a further limit of the spacetime correlators  we find this curve to be the unit circle.  In this case, we also check that the entire worldsheet correlator displays the appropriate crossing symmetry. The non-renormalization of the extremal correlators in the 't Hooft coupling offers a potential
window for a comparison of these results with those from strong coupling.}

\begin{document}
\baselineskip 3.5ex

\section{Introduction}

Since the seminal work of 't Hooft \cite{'tHooft:1973jz} it has been hoped that 
a general gauge theory can be reinterpreted as a string theory with the string coupling $g_s$ 
proportional to the inverse of the rank of the gauge group. For instance, in the case of $SU(N)$, $g_s \sim 1/N$. 
When the gauge theory has a large amount of supersymmetry and is nearly conformal there is a plausible candidate dual in terms of conventional type IIB superstrings propagating on a ten dimensional target space. When the gauge theory is strongly coupled this target space is a geometrical background 
which is asymptotically of the form $AdS_5 \times X$, where  $X$ is
an appropriate five dimensional Sasaki-Einstein manifold. The simplest member of this class being of course $AdS_5 \times S^5$ describing $\mathcal{N} = 4$ SYM theory \cite{Maldacena:1997re}.

On the other hand when the gauge theory is weakly coupled, the dual string theory is complicated and
need not have a simple geometric interpretation. In this paper we will pursue a particular  
proposal \cite{Gopakumar:2003ns,Gopakumar:2004qb,Gopakumar:2004ys,Gopakumar:2005fx} 
 to construct the  worldsheet correlators of such string theories.
This proposal is based on a direct map from the Feynman graphs to the worldsheet moduli, so that one can rewrite (a whole family of) Feynman integrals as an integral
over the appropriate closed string moduli space, i.e. the moduli space of Riemann surfaces with punctures. The essential ingredient in this mapping involves expressing the Feynman amplitude in Schwinger parametrised form and mapping the Schwinger parameters to the Strebel parametrisation of 
the closed string moduli space.  
We will briefly review the salient features of this proposal in what follows. 

Various properties of the correlators obtained by this mapping have been studied in 
\cite{Aharony:2006th,GopakumarDavid,AharonyGopakumarDavid,
Akhmedov:2004yb,Carfora:2006nj,Carfora:2007hj,Furuuchi:2005qm,Yaakov:2006ce,Razamat:2008zr}. Since the first non-trivial instance of closed string moduli arise for the four point functions, much 
of the focus has been on studying particular gauge theory four point functions in the free limit.  
In this context a natural class of correlators to examine are those which are independent of the 'thooft 
coupling. The  "extremal correlators" in  $\mathcal{N}=4$ SYM are an instance where the supersymmetry protects the value of these correlators from any renormalisation \cite{Lee:1998bxa,D'Hoker:1999ea}. We will therefore examine
four point correlators such as $\langle{\rm{Tr}}(Z^{J_1}(x)){\rm{Tr}}(Z^{J_2}(y)){\rm{Tr}}(Z^{J_3}(z))
{\rm{Tr}} (\bar{Z}^{J}(0)))\rangle$, with $J = J_1 + J_2 + J_3$ (Here $Z$ is one of the three complex scalars and using translation invariance we have put the point of insertion of one of the operators at the origin).
The gauge theory spacetime correlators are related by the AdS/CFT correspondence to
string theory worldsheet correlators, integrated over the appropriate 
moduli space, in this case parametrised by the worldsheet cross ratio. If the former do not receive corrections from their free field value, it is interesting to see whether whether the latter do. It could very well be that any coupling dependence can only be given by a total differential on moduli space which would be a BRST trivial addition to the correlator.  In which case the results obtained from any prescription for obtaining the worldsheet answer at zero coupling could be extrapolated to strong coupling and perhaps be compared with a semiclassical worldsheet computation. 
Note that the ${\rm {Tr}}(Z^{J})$ are chiral primary 
operators of $\mathcal{N} = 4$ SYM theory and in the dual string theory correspond to KK modes on $S^5$ with angular momentum $J$. 

\begin{figure}[htbp]
\begin{center}
\epsfig{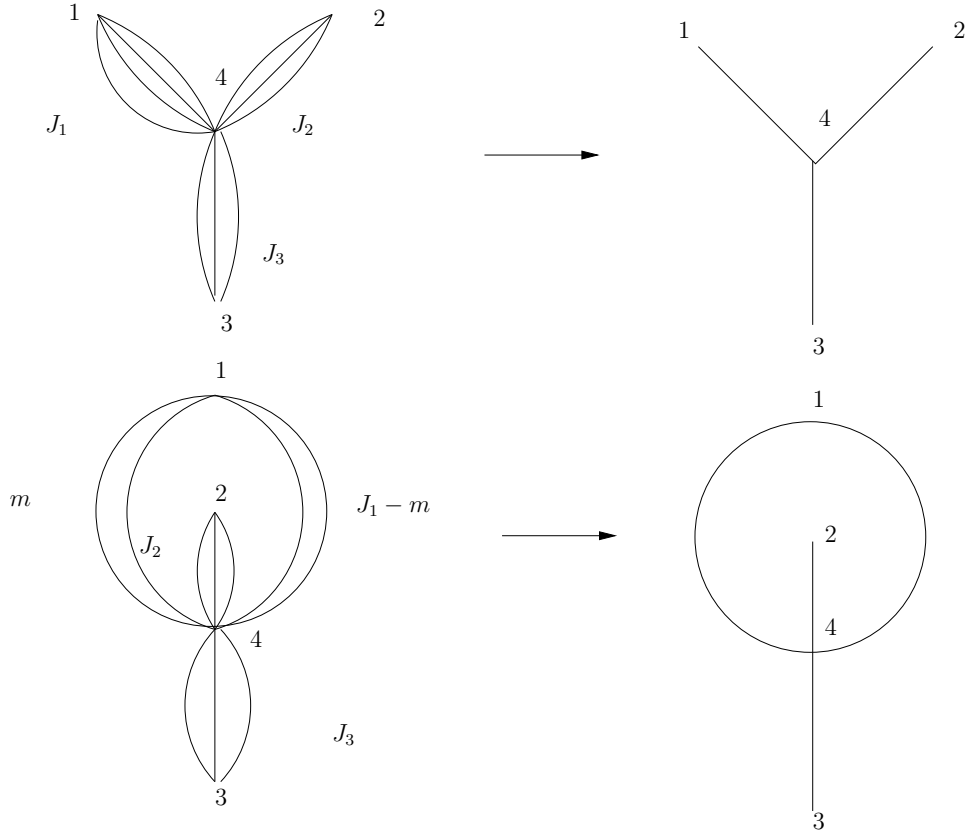} \caption{Here we have shown how we obtain the Y and lollipop skeleton graphs after gluing all the homotopic edges between each pair of vertices. We have suppressed the doubled line notation but these should be thought of as planar doubled lined graphs with homotopic edges glued together. Note we have a family of lollipop graphs, each being distinguished by $m$, which tells us the number of homotopic edges being glued on each of the inequivalent edges joining vertices 1 and 4. The vertex 1 corresponds to the operator insertion ${\rm{Tr}}(Z^{J_1})$ at $x$, vertex 2 to the insertion of ${\rm{Tr}}(Z^{J_2})$ at $y$, vertex 3 to the insertion of 
${\rm {Tr}}(Z^{J_3})$ at $z$ and finally vertex 4 to the insertion of ${\rm{ Tr}} (\bar{Z}^{J})$ at the origin.}\label{yloll}
\end{center}
\end{figure}

With this motivation, we will study the planar Feynman diagrams in the free theory which contribute to 
the above four point extremal correlator. Up to homotopy there are only two classes of diagrams we have to take care of, which we will denote by the Y and the lollipop 
(see  fig.\ref{yloll}). While the Y diagram
is unique, there is a family of lollipop diagrams, since there are two homotopically inequivalent edges joining a pair of vertices and
each member of the family is distinguished by the number of lines glued together in each edge. We will henceforth only draw these skeleton graphs in which homotopically inequivalent edges have been glued together.

The worldsheet correlator for the Y diagram has been evaluated earlier 
\cite{Aharony:2006th,GopakumarDavid}. It was found \cite{GopakumarDavid} that
the worldsheet correlator was a rational function of the crossratio, displayed crossing symmetry and also had an integral power series expansion, all properties required of a correlator in a local CFT. Moreover, the answer was essentially built from Ising model correlators. 
Thus to obtain the entire planar contribution to the worldsheet extremal correlator, we only need to evaluate the lollipop diagrams. However, it appears not to be feasible to solve the lollipop contribution for the most general case, i.e for general values of $J_1, J_2, J_3$. We will study the case where the $J_i \gg 1$. \footnote{Note that this is different from the BMN limit where the mass dimension $J$ is scaled together with $N$.} The advantage of taking this limit is  that we can use a novel saddle point method in the integral over moduli space \cite{AharonyGopakumarDavid},
to obtain the dominant contribution to the correlator. In the dual string theory this limit corresponds to the semiclassical scattering of high energy KK modes in the gravity multiplet. 

We therefore use the saddle point method to find the contribution of the worldsheet correlator from the whole family of lollipop diagrams. This saddle point contribution is exact in our limit of infinite mass dimensions of the operators with their ratios kept fixed. Several novel features  emerge from our analysis. We find that each lollipop diagram gets a contribution supported on a specfic point in the moduli space, i.e for a specific value of the cross ratio. The contributions from the entire family of diagrams then span out a curve in the moduli space. The curve itself parametrically depends  on two combinations of the ratios of the positions as well as the $J_i$. We may contrast this feature with high energy string scattering in flat space as analyzed by Gross and Mende earlier \cite{Gross:1987ar,Gross:1987kza}. Here, it was found that the contribution came from a single point in the moduli space or from a specific value of the cross ratio. We implicitly find this curve and its contribution to the worldsheet correlator for any value of the two independent variables.  However we can write down an explicit answer when both the variables take the value one, which is the case when $x^2/ J_1 = y^2/J_2 = z^2/J_3$, In this case we find that our saddle line (on which the worldsheet correlator is supported) is the unit circle.  We also find that our world sheet correlator is a rational function of the crossratio. 

Another interesting feature of the worldsheet correlator in this limit has to do with the way crossing symmetry is manifested. Crossing symmetry has been a useful consistency check of the worldsheet theory and has played an important role in the case of the Y diagram \cite{GopakumarDavid}. Crossing symmetry is expected  since the lollipop graph is invariant under simultaneous interchange of vertices 2 and 3 and the operators amounting to simultaneous interchange of $y, z$ and $J_2, J_3$ (see fig \ref{yloll}). So we expect that the functional dependence of the worldsheet correlator on the cross ratio should also not change aside from an appropriate weight factor under the transformation of the crossratio which exchanges the point of insertion of vertices 2 and 3 on the worldsheet keeping the other vertices fixed along with the simultaneous exchange of $y, z$ and $J_2, J_3$. We find that this is indeed the case for the explicit example which we have evaluated. The interesting feature is that we need to sum up the contributions from the whole family of lollipop diagrams to see this crossing symmetry.

This paper is organised as follows. In section 2, after briefly recapitulating  the proposal of 
\cite{Gopakumar:2003ns,Gopakumar:2004qb,Gopakumar:2004ys,Gopakumar:2005fx} we go onto study the Strebel differential needed to map the 
Schwinger parameters of the lollipop graph to the worldsheet crossratio. In section 3 we take the large $J_i$ limit. We find, after summing the contributions of these diagrams,  the support of the correlator to lie on a curve in the moduli space. In section 4 we find the explicit answer for the further special case  and we explicitly check crossing symmetry. Finally we conclude with some discussions on the implications of our work for high energy string scattering in AdS.

\section{The Lollipop Diagram}

In this section we introduce the lollipop diagram and write down
the equations which determine the map from the 
space of Schwinger parameters of the lollipop to the 
moduli space of the four punctured sphere. In the course of doing this we will review the 
precise formulation \cite{Gopakumar:2003ns,Gopakumar:2004qb,Gopakumar:2004ys,Gopakumar:2005fx} of this map.

\subsection{The Prescription for the Strebel Differential}

From fig. 1 we see that the lollipop diagram  contributes to 
the planar part of the four point extremal correlator
\begin{equation}
\label{mcorr}
<{\rm Tr}(Z^{J_1}(x)){\rm Tr}(Z^{J_2}(y)){\rm Tr}(Z^{J_3}(z)){\rm Tr}(\bar{Z}^{J}(0)))>,
\end{equation}
where $J = J^1 + J^2 + J^3$. 
To make the discussion self contained we review the 
proposal   \cite{Gopakumar:2003ns,Gopakumar:2004qb,Gopakumar:2004ys,Gopakumar:2005fx}, of   constructing the worldsheet correlator from the 
Schwinger parameterization of the corresponding field theory diagram focussing on 
the correlator in (\ref{mcorr}) as the example.
The proposal of \cite{Gopakumar:2003ns,Gopakumar:2004qb,Gopakumar:2004ys,Gopakumar:2005fx}  involves four steps:

\vspace {.5cm}
\noindent
\emph{(i) Gluing homotopic edges.}  
\vspace{.5cm}

We first simplify the planar\footnote{Here we  focus on only the planar part of the Feynman graph, but 
the proposal of \cite{Gopakumar:2003ns,Gopakumar:2004qb,Gopakumar:2004ys,Gopakumar:2005fx} can be implemented on any genus $g$
Feynman graph.}
 Feynman graph by gluing homotopic edges to a skeleton
graph. 
Two edges joining a pair of vertices are said to be homotopic if they could be made to coincide with the same orientation without any obstruction on the sphere. Mathematically the gluing procedure means that for all the $K$ homotopic edges joining a pair of vertices we use a single effective Schwinger parameter employing the simple identity $$\int d\sigma \sigma^{K-1} exp(-\sigma x^2) \sim 1/ (x^{2})^{K}.$$

For the extremal correlator in (\ref{mcorr}), 
the only allowed Wick contractions are 
 between $Z$ and $\bar{Z}$. Therefore the only edges of the 
 Feynman graph are those 
 between  the origin at which the operator  ${\rm Tr} (\bar{Z}^{J})$
is located   and the points $x$, $y$ and $z$ at which the operators 
${\rm Tr} (\bar{Z}^{J_1})$, ${\rm Tr} (\bar{Z}^{J_2})$, ${\rm Tr} (\bar{Z}^{J_3})$ are located respectively (see fig.\ref{yloll}).
 As mentioned in the introduction we see that  there are two kinds of diagrams, the Y and the lollipop. Furthermore,  from fig.\ref{yloll},  we see that there is in fact a family of lollipop diagrams, each member distinguished by the number of contractions   between the vertices at 1 and  4.  On one of the edges there are $m$ and in the other we have $J_1 - m$ contractions since  the total number of contractions between $x$
 at which vertex $1$ is located  and the origin
 at which vertex $4$ is located  is $J_1$. The vertices 2 and 3 are  at $y$ and $z$ respectively and for all diagrams we have $J_2$ and $J_3$ contractions with the origin respectively. 
 From this discussion it is clear that  the Y is a special case of the lollipop diagram with $m = 0$, or $J_1$. 
 We can always permute $x$, $y$ and $z$ to produce inequivalent graphs, 
 though in some cases the result of permutation would reproduce the original graph. 
 We will discuss the effect of permutations in some detail in section 4.1.
 The left hand side diagrams in fig.1 are the corresponding skeleton graphs
 of the Y and the lollipop diagrams.
 
The Schwinger parameter representation of the whole family of lollipop graphs after all the contractions have been performed is given by
\begin{eqnarray}\label{wick}
& & \sum_{m=1}^{J_1-1} \frac{1}{(J_{1}-m-1)!(m-1)!(J_2-1)!(J_3-1)!}\int_{0}^{\infty}\int_{0}^{\infty}\int_{0}^{\infty}\int_{0}^{\infty} d\sigma_1^{'} d\sigma_1^{''}d\sigma_{2}d\sigma_{3}  \nonumber \\
& &\;\;\;\;\;\;\;\;\;\;\; \times \sigma_{1}^{'m-1}\sigma_{1}^{''J_{1}-m-1}\sigma_{2}^{J_{2}-1}\sigma_{3}^{J_{3}-1}e^{-(\sigma_{1}^{'}+\sigma_{1}^{''})x^{2}-\sigma_{2}y^{2}-\sigma_{3}z^{2}}.
\end{eqnarray}
When 
$m$=0 (or equivalently $m = J_{1}$) the above Schwinger parameterization reduces to
that of the $Y$ diagram. 

\vspace{.5cm}
\noindent
\emph{(ii). The skeleton graph and the critical graph of a Strebel differential} 

\begin{figure}[htbp]
\begin{center}
\epsfig{file = 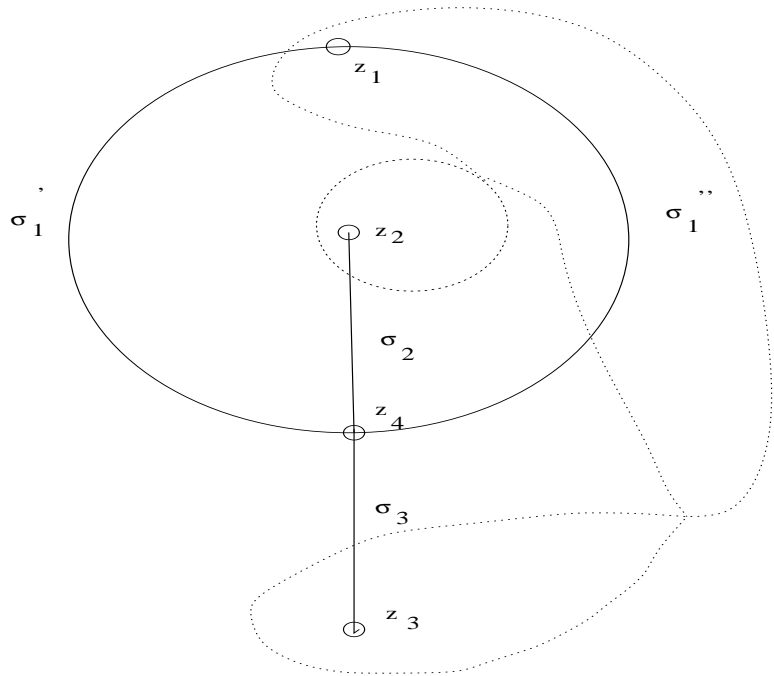, width=4.0in, height = 3.5in}\caption{The dual of the lollipop graph is shown above in dotted lines. The lollipop has two faces, so the dual graph has two vertices. We find that each of these vertices have valence four, which means that the corresponding Strebel differential will have two double zeroes.}\label{lolldual}
\end{center}
\end{figure}

Strebel differentials are special quadratic differentials $\phi(z) dz^2$ whose only poles are double poles and are such that the invariant line element $\sqrt{\phi(z)}dz$ is real on finitely few trajectories. These trajectories are called horizontal trajectories and they connect the zeroes of the Strebel differential. We use Strebel differentials because we can specify a Strebel differential uniquely in two ways, one by specifying the critical graph and the other by specifying the punctured genus $g$ surface (i.e a point in $\mathcal{M}_{g,n}$) along with the residues at the poles (which are a set of $n$ nonnegative numbers specifying a point in $\mathcal{R}_+^{n}$). So we see that it provides the natural tool for mapping the space of graphs to the decorated moduli space of punctured Reimann surfaces $\mathcal{M}_{g,n} \times R_{+}^{n}$. 
The proposal of \cite{Gopakumar:2003ns,Gopakumar:2004qb,Gopakumar:2004ys,Gopakumar:2005fx} 
is to identify the dual of the field theory graph (the lollipop) with the critical graph of an appropriate Strebel differential. 
The Y graph has been already studied in \cite{Aharony:2006th,GopakumarDavid}, we 
therefore focus  on the lollipop graphs. The dual of the lollipop graph is constructed in fig. \ref{lolldual}, we find that it has two vertices, of valence four each and has a bead like structure.

\begin{figure}[htbp]
\begin{center}
\epsfig{file = 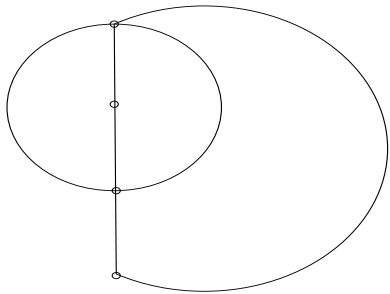, width=2.0in, height = 1.5in}\caption{The maximal completion of the lollipop graph with six edges is shown above. Note if we add any extra edge it would be homotopic to a pre-existing one.}\label{max}
\end{center}
\end{figure}

\begin{figure}[htbp]
\begin{center}
\epsfig{file = 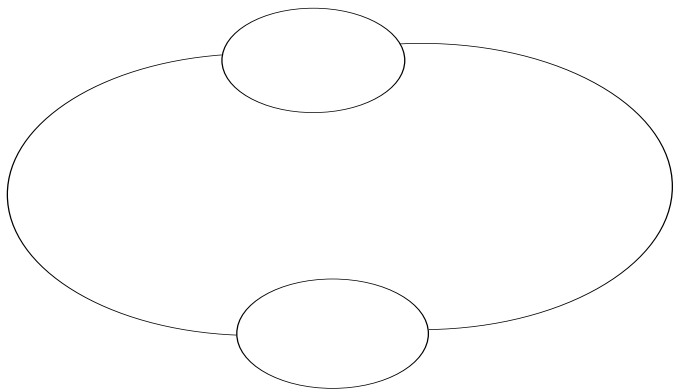, width=3.5in, height = 2.0in}\caption{The dual of the maximally connected graph in fig.3.}\label{maxdual}
\end{center}
\end{figure}

\begin{figure}[h]
\begin{center}
\epsfig{file = 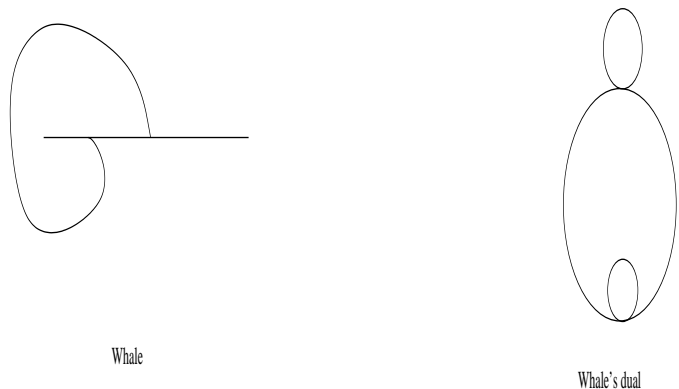, width=3.5in, height = 2.0in}\caption{The whale graph and its dual are shown above. We have either the dual graph of the whale or that of the lollipop as the critical graph of a Strebel differential, depending on the values of the residues of the poles.}\label{whale}
\end{center}
\end{figure}

We now have to identify the dual of the lollipop to the critical graph of the 
appropriate Strebel differential.
For this it is useful   to study more general four vertex graphs on the sphere and
their corresponding Strebel differentials. This will also give us the opportunity to recall the 
properties of Strebel differentials.
The skeleton graph
corresponding to a general four vertex graph will have six edges. For the Lollipop, one such graph
can be obtained from the 
by adding two edges as in fig.\ref{max}. 
Since the graph is actually on a sphere, it provides a triangulation of the sphere and 
we refer to it as a pyramidal tetrahedron. 
This graph is shown in fig.\ref{max} and its dual has been shown in \ref{maxdual}.
We will show below that we can map this tetrahedron  to a cell in the moduli space of the Riemann sphere with four punctures. All such tetrahedra together will provide a cellular decomposition of the moduli space. 
Now the dual of each such tetrahedron is identified with the
 the critical graph of the most general Strebel differential on the sphere with four double poles, 
this differential is given by
\begin{equation}\label{genstrebel}
\phi(z)dz^{2} = -C\frac{(z^{2}-1)(z^{2}k^{2}-1)}{(z-z_{1})^{2}(z-z_{2})^{2}(z-z_{3})^{2}(z-z_{4})^{2}}dz^{2}.
\end{equation}
Here we briefly summarize its properties:
\begin{enumerate}
\item
  The four double poles correspond to the four vertices of the skeleton graph and will be identified with
 closed string vertex operator insertions. 
 Each face of the critical graph associated with this 
 differential encloses the double poles of this differential.
\item
In (\ref{genstrebel}) we have used $SL(2, C)$ transformations to bring the 
four single zeros at $1,-1,1/k$ and $-1/k$. 
The vertices of the critical graph are the zeros of the differential (\ref{genstrebel}), with 
a vertex of valence $m+2$ associated with a zero of order $m$. 
Therefore each of 
the four vertices of the critical graph  of the  differential (\ref{genstrebel})
has $3$ edges.
\item
 On the critical graph, the line element $\sqrt{\phi(z)}dz$ is real everywhere. There are six edges in the critical graph for generic values of the twelve real parameters (which are the six complex numbers $C, k, z_{1}, z_{2}, z_{3}, z_{4}$).
\end{enumerate}
It appears from property 3, we need $12$ real parameters to specify the 
Strebel differential, but we now show that we just need $6$. 
Let us first 
fix the residues of the four poles to be $p_{1}, p_{2}, p_{3}, p_{4}$.
These are real by property 3, thus fixing the 
residues gives us  eight real equations, we can also choose them
to be positive by choosing the directions of the contours. Now the sum 
 of the edges in every face is the residue of the pole inside therefore we have $6-4=2$ independent edges. 
 The lengths of these edges is real by property 3 which implies
that the imaginary parts of the integrals $\sqrt{\phi(z)}dz$ along the edges vanish.
This gives us us two more conditions from the $2$ independent lengths, hence we have ten conditions in all.
So the initial twelve parameters get constrained by ten conditions therefore we actually have two real or one complex parameter left in the end, which we can choose to be the cross ratio $\eta$ of the location of the four poles ($z_{1},z_{2},z_{3},z_{4}$).
Strebel's theorem says that corresponding to six real parameters ($p_{1}, p_{2}, p_{3}, p_{4}, \eta$) we have a unique quadratic differential satisfying all the above properties.
The crucial point is that the number of edges of the original graph equals the number of parameters characterizing a Strebel differential. Thus specifying the length of the dual edges which we refer as 
Strebel lengths amounts to picking a point in the decorated moduli space of a Riemann sphere with four punctures, which specifies the cross ratio and the four real and nonnegative residues at the double poles.
To obtain the Strebel differential for the lolipop note that 
the dual of the lollipop has 2 vertices each of valency 4. This implies that the
corresponding Strebel differential has $2$ zeros each of order $2$. Thus, the 
Strebel differential for the lollipop is given by
\begin{equation}
\label{strebel}
\phi(z)dz^{2} = -C\frac{z^{2}}{(z-z_{1})^{2}(z-z_{2})^{2}(z-z_{3})^{2}(z-z_{4})^{2}}dz^{2}.
\end{equation}
It is clear that the above differential has a double zero at the origin, but by
applying the transformation $z\rightarrow 1/z$ it is easy to verify that the 
differential also has a zero at $\infty$. In the next section we will describe 
 how to obtain the above differential from the most Strebel differential, for the 
 four punctured sphere in (\ref{genstrebel}), using a scaling limit.
Note that there is another graph with four edges and two vertices each of of valency four, hence described by the same Strebel differential. However what 
distinguishes  the lollipop is that the  sum of the residues of the three poles equals that of the fourth pole. 
This relation between the poles can be easily seen  from fig \ref{lolldual}. 
In this paper,  we  will always impose this  relation between the residues of the poles to choose the region of the decorated moduli space which corresponds to the lollipop graph. For the sake of completeness we have drawn the other graph which is
referred as the whale and its dual in fig. \ref{whale}.

\vspace{.5cm}
\noindent
\emph{(iii).  Mapping  Schwinger parameters to the moduli of the
four punctured  sphere}
\vspace{.5cm}

This is the key step of the proposal \cite{Gopakumar:2003ns,Gopakumar:2004qb,Gopakumar:2004ys,Gopakumar:2005fx}.
Let, the Strebel length of the edge of the
  critical graph of the Strebel differential, running between the i-th and j-th zeroes be $l_{ij}$. This is a function of the $p$'s and the crossratio $\eta$. 
 This length
 is identified with  the Schwinger parameter of the \emph{dual} edge in the field theory graph, which we will call $\sigma_{ij}$ and is given by
 $$\sigma_{ij} = l_{ij}(p_1, p_2, p_3, p_4, \eta).$$
 If the graph is not maximally complete the $p$'s and the crossratio would not be independent of each other. However by Strebel's theorem this map would always specify the Strebel differential and hence a point in the decorated moduli space uniquely. 
 We use this map to make a change of variables of the Schwinger integrand from
 $\sigma_{ij}$ to the moduli $\{p_1, p_2, p_3, p_4, \eta\}$.

We now discuss the relations between the Strebel lengths for the lollipop.
From the fig. \ref{lolldual},  it is clear that we have the following relation between the 
residues at the four poles.
\begin{equation}
\label{relres}
p_4 = p_1+p_2 +p_3.
\end{equation}
Let $\sigma_1',\sigma_1^{\prime\prime},  \sigma_2, \sigma_3$ be the 
Strebel lengths as defined in the fig. \ref{lolldual}. We then have the 
following relations:
\begin{eqnarray}
\label{plrelations}
\sigma_{1}^{'}+\sigma_{1}^{''}= p_{1},\\\nonumber
\sigma_{2}=p_{2},\\\nonumber
\sigma_{3}=p_{3}.
\end{eqnarray}
From the above relations, we see that  the lollipop has only one independent Strebel length which we will choose to be $b = \sigma_{1}^{'}$.
Since the Strebel length $b$ is real, it provides a single real condition on the 
Strebel differential in (\ref{strebel}). 
Including the $8$ real conditions determining the residues of the 
poles we have totally $9$ conditions. Out of the 
$10$ real parameters $\{C, z_{1}, z_{2}, z_{3}, z_{4}\}$ in 
 (\ref{strebel}) determining the 
Strebel differential for the lollipop, it is sufficient to specify the 
$4$ real parameters.
These are the  
$3$ independent residues, say $p_1, p_2, p_3$ and one more real parameter.
We can choose the parameters as $p_{2}/p_{1}, p_{3}/p_{1}, \eta$, 
where $\eta$ is the cross ratio of the location of the four poles.
Note that the number of parameters equals the number of edges, therefore
specifying the Strebel lengths amounts to choosing a point in the moduli
space of the Riemann sphere with four punctures.
 
\vspace{.5cm}
\noindent
\emph{(iv).  Integrating the residues}
\vspace{.5cm}

The final step in of the proposal \cite{Gopakumar:2003ns,Gopakumar:2004qb,Gopakumar:2004ys,Gopakumar:2005fx} is to integrate the 
independent residues and identity the integrand as the putative world sheet 
correlator. For the lollipop diagram, 
after the change of variables from the Schwinger parameters  to the  ratios
$p_2/p_1, p_3/p_2, \eta $, we have to 
integrate over the variables 
$p_2/p_1, p_3/p_2$ to obtain an integrand
which depends only on $\eta$. Thus we are left with
$$
\int d\eta d\bar{\eta} {\cal G}(\eta, \bar\eta).
$$
We then identify ${\cal G}(\eta, \bar\eta)$ as the candidate world sheet correlaor.

\subsection{The Strebel Differential for the Lollipop}

In this subsection we write down the equations which determine the 
map from the Schwinger parameters to the moduli of the four punctured 
sphere.
To do this we will first solve the independent parameters of the Strebel differential (\ref{strebel}), which are two of the four residues of the double poles and the complex crossratio $\eta$ in terms of the lengths of the four edges of the critical graph.
The four edges of the critical graph are 
 now identified with the  Schwinger parameters of the dual edges. 

As stated in the previous section it would be useful to begin with the most general Strebel differential whose critical graph (fig.\ref{maxdual}) is maximally complete with six edges. 
We will then 
take an appropriate scaling limit that would take us to the special case of the lollipop graph. 
We follow the approach developed in \cite{GopakumarDavid}.
First we 
map the general Strebel differential (\ref{genstrebel}) to an auxiliary torus with complex coordinate $u$. 
In the $u$ plane the doubly periodic properties of the the differential
$\sqrt{\phi(z)} dz$ are more manifest.
This torus has periods ($2\omega_1, 2 \omega_2)$) 
which are functions of the
moduls $k$.
The explicit map (from $z \rightarrow u(z)$) is given by:
\begin{equation}
u(z) =\int_{1}^{z}\frac{dz}{\sqrt{(z^{2}-1)(z^{2}k^{2}-1)}}.
\end{equation}
Upto a constant shift this is defining relation for the Jacobi elliptic function 
$\sn(u)$ of modulus $k$. We have the relation
\begin{equation}
\label{relz}
z= \sn\left( u+ \frac{\omega_1}{2} \right) = \frac{cn(u)}{\dn(u)}.
\end{equation}
In the $u$-plane the invariant line element is given by
\begin{equation}
\sqrt{\phi(z)} dz = -i\sqrt{C}(1-k^{2})^{2}\frac{sn^{2}(u)}{\Pi_{i=1}^{4}(cn(u)-z_{i}dn(u))}du.
\end{equation}
Note that in the $u$-plane we have no branch cuts therefore the invariant line element is single valued in the $u$-plane. This line element has double zeroes at $0, \omega_1,
\omega_2, \omega_1 +\omega_2$. 
It has single poles at $z_i = \cn(u_i)/\dn(u_i)$, it can be shown \cite{GopakumarDavid}
that the residues at these poles $r_i$ satisfy the following equations
\begin{eqnarray}
\label{basiceq}
\sum_{i=1}^4\frac{r_i}{\sn(u_i)}=0, \\\nonumber
\sum_{i=1}^4 r_{i} \sn(u_{i}) = 0, \\\nonumber
\sum_{i=1}^4 r_{i} \frac{\cn(u_{i})\dn(u_{i})}{\sn(u_{i})} =0 .
\end{eqnarray}
Here $z_i =\cn(u_i)/\dn(u_i)$. In principle the equations (\ref{basiceq}) 
determine three of 
the four $u_i$ in terms of the fourth, say $u_1$ as well as the perimeters $r_i$. 
to determine $u_0$ we need to know the two Strebel lengths. 
The main reason to go to the $u$-plane is that it enables us to carry out the 
Strebel integrals between the zeros. The two independent lengths can be
taken to be 
\begin{equation}
a = \int_0^{\omega_1} \sqrt{\phi(u)} du, 
\qquad \hbox{and} \qquad 
b = - \int_0^{\omega_2} \sqrt{\phi(u)} du.
\end{equation}
After performing these integrals we obtain \cite{GopakumarDavid}
\begin{eqnarray}
\label{defab}
a = \sum_i r_i \left[ \pi - 2i ( \zeta(u_i) \omega_1 - \zeta(\omega_1) u_i) \right],
\\ \nonumber
b = \sum_i r_i \left[ \pi + 2i ( \zeta(u_i) \omega_2 - \zeta(\omega_2) u_i) \right].
\end{eqnarray}
where $\zeta(u)$ is related to the Weierstrass ${\wp}$ function by
\footnote{Please refer \cite{GopakumarDavid} for a discussion on the
properties of these functions.}
\begin{equation}
\frac{d\zeta}{dz} = -{\wp}(z).
\end{equation}
By taking linear combinations of the equations in (\ref{defab}) and using the 
properties of these functions we obtain 
\begin{equation}
\label{niccomb}
\pi \sum_i r_i u_i = 
(\pi \omega_1 + \pi \omega_2) \sum_i r_i - a \omega_2 - b\omega_1.
\end{equation}

We now need to take a scaling limit of the most general Strebel differential
for the four punctured sphere given in \eq{genstrebel} so as to obtain the Strebel
differential for the lollipop given in \eq{strebel}. This scaling limit is given by
\begin{eqnarray}
\label{scallimit}
z\rightarrow z'/\epsilon, \qquad z_i \rightarrow z_i'/\epsilon,  \\ \nonumber
k \rightarrow k' \epsilon^{1+\alpha}, \qquad 
{\hbox{with}}\;\; \epsilon \rightarrow 0.
\end{eqnarray}
where $1>\alpha >0$, 
Note that $z', z_i', k'$ are held fixed. Under this scaling the zeros of the 
Strebel differential are pushed to $\pm\epsilon, \pm 1/\epsilon^{2\alpha}$. 
Therefore in the limit we see that the general Strebel differential in \eq{genstrebel}
coincides with that of the lollipop \eq{strebel}. 
We now examine how the equations \eq{basiceq} behave under this scaling.
From \eq{relz} we can obtain the following expansions under the scaling \eq{scallimit}
\begin{eqnarray}
\label{expansion}
\sn u = i\frac{z'}{\epsilon} \left( 
1+ \frac{1}{2} k^{\prime 2} \epsilon^{2\alpha} z^{\prime 2} + \cdots \right) , \\ \nonumber
\frac{1}{\sn u} = -i \frac{\epsilon}{z} \left( 1 - 
\frac{1}{2} k^{\prime 2} \epsilon^{2\alpha} z^{\prime 2} + \cdots \right), \\ \nonumber
\frac{\cn u \dn u}{\sn u} = i \left( 1 + 
 \frac{1}{2} k^{\prime 2} \epsilon^{2\alpha} z^{\prime 2} + \cdots \right).
 \eea
 To perform this expansion we have used $0< 2\alpha<1$, the $\cdots$ refer to 
 higher powers of $\epsilon$. 
 Substituting these equations in \eq{basiceq}, the leading order equations are 
 given by
 \begin{equation}
 \label{leadeq}
 \sum \frac{r_i}{z_i}=0,\qquad
\sum r_{i}z_{i}=0,\qquad
\sum r_{i}=0. 
\end{equation}
We now have to take the scaling limit  in the equations determining the 
Strebel lengths \eq{defab}, \eq{niccomb}. We will first look at the equations
determining $a$. It can be shown that under the scaling limit in \eq{scallimit}
\begin{eqnarray}
\label{scalzeta}
iu &=&  \ln\left( \frac{2z'}{\epsilon}\right) + \frac{z'}{4} \epsilon^{2\alpha}, \\ \nonumber
\zeta(u) &=& -\frac{i}{12 }\ln( \frac{2z'}{\epsilon} )+  \frac{1}{2} + 
 O(\epsilon^{2\alpha}), \\ \nonumber
 \zeta(\omega_1) &=& \frac{\pi}{12}  +  O(\epsilon^{2\alpha}), \\ \nonumber
 \omega_1 &=& \pi +  O(\epsilon^{2\alpha}).
 \end{eqnarray}
 Substituting the leading terms of these expansions for the equation for 
 $a$ in \eq{defab} and using the last equation of
 \eq{leadeq} we obtain that $a=0$ to $O(\epsilon^{2\alpha})$.
 To obtain the equation for $b$ it is 
 convenient to look at \eq{niccomb}. Again substituting the leading terms 
 of \eq{scalzeta} and using $a=0$  we obtain
 \begin{equation}
 \label{simpb}
 \sum_i r_i \ln (z_i) = i b,  \qquad {\hbox{or}} \;\;\;\; 
 z_{1}^{r_1}z_{2}^{r_2}z_{3}^{r_3}z_{4}^{r_4} = e^{ib}.
 \end{equation}
 Here again we have used the last equation in \eq{leadeq}. 
We now further simply the equations \eq{leadeq} and \eq{simpb}. We can eliminate
$r_4$ from the equations by $r_4 = -r_1 -r_2 -r_3$. Let us also define 
\begin{equation}
s_2 = \frac{r_2 }{r_1}, \qquad s_3 = \frac{r_3}{r_1}, \qquad s = \frac{b}{r_1},
\end{equation}
and 
\begin{equation}
v_1 = \frac{z_1 }{z_4}, \qquad v_2 = \frac{z_1 }{z_4}, \qquad v_3 = \frac{z_3 }{z_4}. 
\end{equation}
Then the equations (\ref{leadeq}) and (\ref{simpb}) reduce to
\begin{eqnarray}\label{set1}
v_1 + s_2 v_2 + s_3 v_3 &=& s_2 +s_3 +1, \\\nonumber
\frac{1}{v_1} +\frac{s_2}{v_2} + \frac{s_3}{v_3} &=&  s_2 +s_3 +1, \\\nonumber
v_{1}v_{2}^{s_2}v_{3}^{s_3}& =& e^{is}.
\end{eqnarray}
The crossratio is :
\begin{equation}\label{set2}
\eta = \frac{(z_2 - z_4)(z_3 - z_1)}{(z_3 - z_4)(z_2 - z_1)} = \frac{(v_2 -1)(v_3 -v_1)}{(v_3 -1)(v_2 -v_1)}. 
\end{equation}
Now using the relations (\ref{plrelations}) between the Strebel lengths and the Schwinger parameters  we have 
\begin{eqnarray}\label{plrelations2}
s=\frac{\sigma_{1}^{'}}{\sigma_{1}}, \qquad s_{2}=\frac{\sigma_{2}}{\sigma_{1}},
\qquad s_{3} = \frac{\sigma_{3}}{\sigma_{1}}.
\end{eqnarray}
where $\sigma_{1} = \sigma_{1}^{'}+\sigma_{1}^{''}$.

Thus the  overall scale is indeed unimportant for the map. The stategy is to take $s_{2}, s_{3}$ and the crossratio $\eta$ as the independent variables parameterising our Strebel differential (\ref{strebel}) or equivalently the decorated worldsheet moduli space. We will see later that this choice is good both for obtaining the high energy behaviour of the worldsheet correlator and also for demonstrating crossing symmetry. So from the three equations of (\ref{set1}) we first solve $v_{1}, v_{2}, v_{3}$ as functions of $s_{2}, s_{3}, s$. We then substitute this in the expression for $\eta$ in (\ref{set2}) and invert s as a function of $s_{2}, s_{3}, \eta$. This would complete our job of solving the map from the Schwinger parameters to the variables of choice in the decorated moduli space of four punctured Riemann surface, which are now $s_2, s_3$ and $\eta$.

\subsection{An exactly solvable point of the lollipop}

The equations in \eq{set1} are transcendental and it is not possible to obtain 
exact solutions for all values of $s_2, s_3, s$. However 
when $s_2=s_3=1$ we show the equations can indeed be solved explicitly.
Substituting these values  in \eq{set1} we obtain
\begin{eqnarray}
v_1 + v_2 +v_3 = -3, \\\nonumber
\frac{1}{v_1}+\frac{1}{v_2}+\frac{1}{v_3}=-3, \\\nonumber
v_1 v_2  v_3 = e^{is}.
\end{eqnarray}
In other words $v_1, v_2, v_3$ are the three roots of the cubic equation 
$$v^3 -3v^2 + 3e^{is} v -e^{is} =0.
$$ 
We will show  that the right choice of the roots  is given by:
\begin{eqnarray}\label{choice}
v_1 = \omega^2 A + \omega B +1,  &\qquad&
v_2 = A + B +1, \\\nonumber
v_3 = \omega A + \omega^2 B +1,  &\qquad& \mbox{where }\; 
A =
\beta^{1/3}(1+\sqrt{1-\beta})^{1/3}, \\ \nonumber
B=\beta^{1/3}(1-\sqrt{1-\beta})^{1/3}, &\qquad& \mbox{and} \; \beta=1-e^{is}.
\end{eqnarray}
Here $\omega$ refers to the cube root of unity. 
With this choice of $v$'s we obtain the crossratio using  (\ref{set2}) 
\begin{equation}\label{saddleline}
\eta = \eta_{o}(s) = -\omega\frac{ 1-\frac{B^2}{A^2}}{1-\omega^{2}\frac{B^2}{A^2}}, 
\quad \hbox{where}\;\;
\frac{B^2}{A^2}=\omega^{-\frac{1}{2}}\left(\tan(\frac{s}{4}) \right)^{\frac{2}{3}}. 
\end{equation}
Note that 
 when $s$ is set to  zero, $\eta$ becomes $-\omega$. 
 This justifies the 
  choice of roots in (\ref{choice}):
  because when $s$ vanishes the lollipop graph reduces to  Y graph. 
 In fact,  the exactly solvable point of the lollipop diagram reduces to the 
 the Y diagram with three equal edges when $s=0$. 
  When evaluating the 
 world sheet correlator  corresponding to the extremal operators there is 
 a contribution from the $Y$ graph when the the number of contractions
  in (\ref{wick})) is either $m=0$ or $m =J_1$. 
  It is necessary to have a continuous and uniform definition of the 
 $\eta$ into order to compare and add the contribution from the Y diagrams. 
 From  \cite{GopakumarDavid}, we see that when all the lengths of the 
 $Y$ diagram are equal, the crossratio reduces to $\eta = -\omega$.
 The choice of the roots in (\ref{choice}) ensures that the 
 crossratio  in (\ref{saddleline}) also reduces to $\eta =-\omega$ when 
 $s=0$.

\section{The large $J$ limit}

We have seen in the previous section the change of variables required to 
obtain the world sheet correlator for the lollipop diagram  involves a transcedental
equation. It is therefore difficult to obtain the worldsheet correlator for the 
general diagram. In \cite{AharonyGopakumarDavid} it was observed that 
 correlators with large $J$ charges simplify. It was shown that 
 the Schwinger parametric representation of the correlator localizes 
 at certain values of the Schwinger parameters.
 In this section we study the lollipop in the limit 
 $J_1, J_2, J_3, \rightarrow \infty$
 with their mutual ratios held fixed. 
 In this limit, the general analysis in \cite{AharonyGopakumarDavid} shows that the contribution to the integrand localises at certain values of the Schwinger parameters.

\subsection{Schwinger parametrization in the large $J$ limit}

In the previous section we have noticed that we can always take an overall scale of the Strebel differential out before we perform
the change of variables to the decorated worldsheet moduli space. 
We chose $\sigma_{1} = \sigma_{1}^{'} +\sigma_{1}^{''}$ in (\ref{plrelations2})
to be the overall scale.
Recasting  the Schwinger parametric representation in terms of the scaled residues $s, s_2$ and $s_3$ as specified in equation (\ref{plrelations2}) we obtain
\begin{eqnarray}
& & \sum_{m=1}^{J_1-1} \frac{(J-1)!}{(J_{1}-m-1)!(m-1)!(J_2-1)!(J_3-1)!}\int_{0}^{1}\int_{0}^{\infty}\int_{0}^{\infty} dsds_{2}ds_{3} \nonumber \\
& &~~~~~~~~~~~~~~~~~~~~~~~~~
\times \frac{s^{m-1}(1-s)^{J_{1}-m-1}s_{2}^{J_{2}-1}s_{3}^{J_{3}-1}}{(x^{2}+s_{2}y^{2}+s_{3}z^{2})^{J_{1}+J_{2}+J_{3}}}.
\end{eqnarray}
In principle we should have mapped each member of the family of lollipop contributions, denoted by 
$m$, to the worldsheet moduli space for each $m$ separately. But, the change of variables  depends just 
 on the topology of the graph  and not on the number of lines glued at each edge, therefore not on $m$. Hence we can  the sum over $m$  first and then perform the further change of variables which would eventually include the crossratio of the location of the four poles of the Strebel differential. The sum over $m$ gives:
\begin{equation}\label{summed}
\frac{(J-1)!}{(J_{1}-2)!(J_2-1)!(J_3-1)!}\int_{0}^{1}ds\int_{0}^{\infty}ds_{2}\int_{0}^{\infty}ds_{3}\frac{s_{2}^{J_{2}-1}s_{3}^{J_{3}-1}}{(x^{2}+s_{2}y^{2}+s_{3}z^{2})^{J}}, 
\end{equation}
where $J =J_1 + J_2 +J_3$.  It is sufficient to focus on the integral to 
discuss the change of variables, therefore for the present we ignore the 
normalization in (\ref{summed}).
To simplify the discussion we scale out
 $d^{2} = x^{2}+y^{2}+z^{2}$ from the denominator. Define $a, b,c$  such that, $x^{2}/d^{2} = a, y^{2}/d^{2} = b, z^{2}/d^{2} = c$. Then the integral becomes
\begin{equation}\label{srepn}
\int_{0}^{1}ds\int_{0}^{\infty}ds_{2}\int_{0}^{\infty}ds_{3}\frac{s_{2}^{J_{2}-1}s_{3}^{J_{3}-1}}{(a+s_{2}b+s_{3}c)^{J}}.
\end{equation}
It will be also convenient to introduce $\alpha = (J_{1}-2)/J, \beta = (J_{2}-1)/J, \gamma=(J_{3}-1)/J$. In the limit of large J, with $\alpha, \beta, \gamma$ held fixed, the saddle point value for $s_{2}$ is $J_{2}a/J_{1}b$, the saddle point for $s_{3}$ is $J_{3}a/J_{1}c$. Note that
  $s$ is a flat direction since the integrand is independent of this variable. We can now expand the integrand about the saddle point. Let us write  $s_{2}=J_2a/J_1b+\epsilon_{2}$ and $s_{3}=J_3a/J_1c+\epsilon_{3}$. 
  Substituting this in the 
  integrand  (\ref{srepn}) we obtain:
\begin{eqnarray}
 && \exp \left[ -J\left(\ln(J)+\alpha \ln(\frac{a}{\alpha}) +\beta \ln(\frac{b}{\beta}) +\gamma \ln(\frac{c}{\gamma}) \right.  \right. \\ \nonumber
&& \;\;\;  \left.\left.  +\ln(1+\frac{b\alpha}{a} \epsilon_{2} + \frac{c\alpha}{a} \epsilon_{3})-\beta \ln(1+\frac{b\alpha}{a\beta}\epsilon_{2}) -\gamma \ln(1+\frac{c\alpha}{a\gamma}\epsilon_{3}) \right) \right] .
\end{eqnarray}
Keeping only upto quadratic terms in the exponent we have:
\begin{eqnarray}\label{saddlexp}
& &\exp\left[ -J\left( \ln(J)+\alpha \ln(\frac{a}{\alpha}) +\beta \ln(\frac{b}{\beta}) +\gamma \ln(\frac{c}{\gamma}) \right. \right. \\ \nonumber
& & \;\;\; \left.\left.  -\frac{b^{2}\alpha^{2}}{2a^{2}}\frac{1-\beta}{\beta}\epsilon_{2}^{2}-\frac{c^{2}\alpha^{2}}{2a^{2}}\frac{1-\gamma}{\gamma}\epsilon_{3}^{2}-\frac{bc\alpha^{2}}{a^{2}}\epsilon_{2}\epsilon_{3}\right) \right] .
\end{eqnarray}
There are no  linear terms in $\epsilon_{2}$ and $\epsilon_{3}$ since
we are expanding about a saddle point. Now in the large $J$ limit the Gaussian fluctuations could be interpreted as delta functions. With the understanding that we are going to do away with $\epsilon_{3}$ integral first, we can write the expression in 
(\ref{saddlexp}) more compactly as
\begin{equation}
e^{-J\left[ \ln(J)+\alpha \ln(\frac{a}{\alpha}) +\beta \ln(\frac{b}{\beta}) +\gamma \ln(\frac{c}{\gamma})\right] }
\left(\frac{2\pi}{J}\right)
\left(\frac{a^{2}
\sqrt{\beta\gamma}}{bc\alpha^{\frac{5}{2}}}\right)
\delta(\epsilon_{2})\delta({\epsilon_3}). 
\end{equation}
Now we note that in all the intermediate stages of manipulations
 the whole expression has retained symmetry under simultaneous exchange between $\beta, b$ and $\gamma, c$. This is the reflection of the fact that 
  the lollipop graph as a double line planar graph has a symmetry under exchange of vertices 2 and 3 (see fig. (\ref{lolldual})).

Aside from a multiplying constant the whole expression is just the integral $$\int ds\int d\epsilon_{2}\int d\epsilon_{3}\delta(\epsilon_{2})\delta({\epsilon_{3}}).$$
 We need to convert this to an integral over $\eta,\bar{\eta}$ , the 
coordinates which parameterize the moduli space of four punctured Reimann sphere and $\epsilon_{3}$. We then have to 
  integrate $\epsilon_{3}$ out to obtain the world sheet correlator,
  this last integration is trivial due to the delta function.

\subsection{The world sheet correlator in the large $J$ limit. }

To obtain the world sheet correlator we
 we have to solve equations (\ref{set1}) and (\ref{set2}) and obtain $v_1, v_2, v_3$
 as functions of $s,s_2,s_3$. In the large $J$ limit it is 
sufficient to solve these equations around the saddle line, i.e, $s_2, s_3$, around their saddle point values $(\beta a)/(\alpha b),(\gamma a)/(\alpha c)$ respectively and the
flat direction $s$ parametrising the saddle line. 
As in the previous subsection, we write $s_2 = (\beta a)/(\alpha b) + \epsilon_2$, $s_3 = (\gamma a)/(\alpha c) + \epsilon_3$. In terms of $\epsilon_{2}$ and $\epsilon_{3}$ 
the solutions to  (\ref{set1})  can be expanded perturbatively as:
\begin{eqnarray}\label{pertsoln}
v_1 = v_{10}(s) + \epsilon_{2}v_{11}(s) +\epsilon_{3}v_{12}(s) + ....., \\\nonumber 
v_2 = v_{20}(s) + \epsilon_{2}v_{21}(s) +\epsilon_{3}v_{22}(s) + ....., \\\nonumber
v_3 = v_{30}(s) + \epsilon_{2}v_{31}(s) +\epsilon_{3}v_{32}(s) + ..... .
\end{eqnarray}
Substituting these expansions  in (\ref{set2}) to obtain $\eta$ , we get:
\begin{equation}
\eta = \eta_{o}(s) + M(s) \epsilon_{2} + N(s) \epsilon_{3}.
\end{equation}
In general $M(s)$ and $N(s)$ are complicated functions, but we will see that 
we will not need their explicit forms to obtain the world sheet correlator.

We have seen that up to some factors which will be 
reinstated later the field theory integrand is given by:
$$\int ds\int d\epsilon_{2}\int d\epsilon_{3}\delta(\epsilon_{2})\delta({\epsilon_{3}}).$$ 
We  now have to  change  variables from $s, \epsilon_{2}, \epsilon_{3}$ to $\eta, \bar{\eta}, \epsilon_{3}$ and then integrate out $\epsilon_{3}$. 
In the large $J$ limit we can write down the following change of variables:
\begin{eqnarray}\label{set3}
\eta &=& \eta_{o}(s) + M(s) \epsilon_{2} + N(s) \epsilon_{3}, \\\nonumber
\bar{\eta} &=& \bar{\eta_{o}}(s)+\bar{M}(s) \epsilon_{2} + \bar{N}(s)\epsilon_{3}, \\\nonumber
\epsilon_{3}& =& \epsilon_3.
\end{eqnarray}

With the understanding that we are integrating out $\epsilon_{3}$ first we can work out the change of variables for the case $\epsilon_{3} \sim 0$ because of the presence of $\delta(\epsilon_{3})$. Note that we also have $\delta(\epsilon_{2})$ in the expression
for the world sheet correlator. Therefore, when we obtain the 
 the Jacobian for $s, \epsilon_{2}$ to $\eta, \bar{\eta}$ we can set
 $\epsilon_{2}$ to zero as well. Now we will think of i$\delta(\epsilon_{2})$ as a 'primary function' of $\bar{\eta}$. This means while doing the integration, we must perform the $\bar{\eta}$ integral first \footnote{If we have a delta function of many variables we need to interpret this distribution as a primary function of one of the variables which we  must integrate first. For example, the familiar $\delta (p^2)$ is a function of four variables, namely the four momenta ($E$, $\textbf{p}$), but we can
  regard it as a primary function of the energy $E$. Therefore we write $\delta(p^2)$ as
  $(\delta(E - \vert \textbf{p} \vert)+ \delta(E + \vert \textbf{p} \vert))/ 2 \vert \textbf{p} \vert.$}. To do this 
   we have to write the Jacobian as a function of $\eta$ only. It is easy to see that:
\begin{equation}\label{Jacobian}
\left. \frac{\partial(s,\epsilon_{2})}{\partial(\eta,\bar{\eta})}
\right|_{\epsilon_{2}=\epsilon_{3}=0}=\frac{1}{M(\eta_{o}^{-1}(\eta))\bar{\eta_{o}^{'}}(\eta_{o}^{-1}(\eta))-\bar{M}(\eta_{o}^{-1}(\eta))\eta_{o}^{'}(\eta_{o}^{-1}(\eta))}.
\end{equation}
 Now, converting  $\delta(\epsilon_{2})$ term, with $\epsilon _{3}=0$  we can write:
\begin{equation}
\delta(\epsilon_{2}) = \delta(\frac{\bar{\eta}-\bar{\eta_{o}}(s(\eta,\bar{\eta}))}{\bar{M}(s(\eta,\bar{\eta}))}).
\end{equation}
Note in converting this delta function we cannot write $s$ as $\eta_{o}^{-1}(\eta)$ like we have done for the Jacobian since $\epsilon_{3} =0$ only outside the Jacobian. However we can still put $\epsilon_{3}$ to zero and then eliminate $\epsilon_{2}$ from (\ref{set3}) to obtain:
\begin{equation}\label{sdefn}
\frac{\bar{\eta}-\bar{\eta_{o}}(s))}{\bar{M}(s)}=\frac{\eta-\eta_{o}(s)}{M(s)} .
\end{equation}
We now use above  expression to write $s$ as a function of $\eta,\bar{\eta}$. Alternatively, we can also Taylor expand $s(\eta,\bar{\eta})$ in $\bar{\eta}$ about $\bar{\eta} = \bar{\eta_{o}}(\eta_{o}^{-1}(\eta))$ which happens when $\epsilon_{2} = 0$. TheTaylor expansion is given by:
\begin{equation}\label{sexpn}
s(\eta,\bar{\eta})=\eta_{o}^{-1}(\eta)+ (\bar{\eta} - \bar{\eta_{o}}(\eta_{o}^{-1}(\eta)))\left. \frac{\partial s(\eta,\bar{\eta})}{\partial \bar{\eta}}\right|_{\bar{\eta} = \bar{\eta_{o}}(\eta_{o}^{-1}(\eta))}+..... .
\end{equation}
We calculate the first derivative $(\partial s(\eta,\bar{\eta}))/(\partial \bar{\eta})$ when $\bar{\eta} = \bar{\eta_{0}}(\eta_{0}^{-1}(\eta))$ using
(\ref{sdefn}),  the defining expression for $s$. We take the derivative on both sides of (\ref{sdefn}) and retain only terms which don't vanish when $\epsilon_{2} = 0$.
This  means we can drop derivatives of $M$ and $\bar{M}$. Thus we obtain :
\begin{equation}\label{sderv}
\frac{\partial s(\eta,\bar{\eta})}{\partial \bar{\eta}}\vert_{\bar{\eta} = \bar{\eta_{o}}(\eta_{o}^{-1}(\eta))}=\frac{M(\eta_{o}^{-1}(\eta))}{M(\eta_{o}^{-1}(\eta))\bar{\eta_{o}^{'}}(\eta_{o}^{-1}(\eta))-\bar{M}(\eta_{o}^{-1}(\eta))\eta_{o}^{'}(\eta_{o}^{-1}(\eta))}.
\end{equation}
Now to finish the required conversion of $\delta(\epsilon_{2})$ to a primary function of $\bar{\eta}$, we see that
\begin{equation}\label{delredn}
\delta\left(\frac{\bar{\eta}-\bar{\eta_{o}}(s(\eta,\bar{\eta}))}{\bar{M}(s(\eta,\bar{\eta}))}\right) = \delta(\bar{\eta} - \bar{\eta_{o}}(\eta_{o}^{-1}(\eta)))\frac{\bar{M}(\eta_{o}^{-1}(\eta))}{1 - \bar{\eta_{o}^{'}}(\eta_{o}^{-1}(\eta))\left.\frac{\partial s(\eta,\bar{\eta})}{\partial \bar{\eta}}\right|_{\bar{\eta} = \bar{\eta_{o}}(\eta_{o}^{-1}(\eta))}} .
\end{equation}
Note we have dropped the derivative of $\bar{M}$ since this is zero when the argument of the delta function vanishes. Now we 
 substitute $\frac{\partial s(\eta,\bar{\eta})}{\partial \bar{\eta}}$ from (\ref{sderv}) in (\ref{delredn})  and obtain:
\begin{eqnarray}\label{delta}
\delta(\epsilon_{2}) &=&\delta\left(\frac{\bar{\eta}-\bar{\eta_{o}}(s(\eta,\bar{\eta}))}{\bar{M}(s(\eta,\bar{\eta}))}\right),   \\\nonumber &=&\delta(\bar{\eta} - \bar{\eta_{o}}(\eta_{o}^{-1}(\eta)))\frac{M(\eta_{o}^{-1}(\eta))\bar{\eta_{o}^{'}}(\eta_{o}^{-1}(\eta))-\bar{M}(\eta_{o}^{-1}(\eta))\eta_{o}^{'}(\eta_{o}^{-1}(\eta))}{\eta_{o}^{'}(\eta_{o}^{-1}(\eta))}.
\end{eqnarray}
Now combining (\ref{delta}) above with the Jacobian (\ref{Jacobian}) we get a remarkably simple answer:
\begin{equation}
\int ds \int d\epsilon_{2} \int d\epsilon_{3} \delta(\epsilon_{2})\delta(\epsilon_{3}) = \int d\eta\int d\bar{\eta}\int d\epsilon_{3}  \frac{\delta(\bar{\eta} - \bar{\eta_{o}}(\eta_{o}^{-1}(\eta)))\delta(\epsilon_{3})}{\eta_{o}^{'}(\eta_{o}^{-1}(\eta))}.
\end{equation}
Now we can integrate $\epsilon_{3}$ out and claim that our correlator is:
\begin{equation}\label{gencorr}
\mathcal{G}(\eta,\bar{\eta}) \sim \frac{\delta(\bar{\eta} - \bar{\eta_{o}}(\eta_{o}^{-1}(\eta)))}{\eta_{o}^{'}(\eta_{o}^{-1}(\eta))} (\hbox{const}).
\end{equation}
This is a very simple answer which could have been expected before performing the detailed calculation. Note the $\bar{\eta}$ integral gives  1 and $d\eta/(\eta_{o}^{'}(\eta_{o}^{-1}(\eta)))$ is simply equal to $ds$. The delta function plays the role of localizing the correlator on the saddle line. 

If we reinstate the constant in (\ref{gencorr}) we get:
\begin{equation}
\frac{(J-1)!}{(J_{1}-2)!(J_2-1)!(J_3-1)!}e^{-J\left[\ln(J)+\alpha \ln(\frac{a}{\alpha}) +\beta \ln(\frac{b}{\beta}) +\gamma \ln(\frac{c}{\gamma})\right)}
\left(\frac{2\pi}{J}\right)\left(\frac{a^{2}\sqrt{\beta\gamma}}{bc\alpha^{\frac{5}{2}}}\right).
\end{equation}
Using Sterling's approximation we can expand the prefactor in the above
expression to obtain:
\begin{equation}\label{constant}
e^{-J\left[\alpha \ln(a) +\beta \ln(b) + \gamma \ln(c)\right]}\left(\frac{2\pi}{J}\right)
\left(\frac{a^{2}\sqrt{\beta\gamma}}{bc\alpha^{\frac{5}{2}}}\right).
\end{equation}
Implicitly (\ref{gencorr}) is the general answer for the world sheet correlator, though we do not know the function $\eta_{o}^{-1}(\eta)$ and its derivative explicitly. 
Note that
we have derived the above expression after we have inserted the operators at the vertices of the lollipop graph in a definite order. We have taken a specfic case where the spacetime point $x$ has been mapped to the vertex 1 of the lollipop as in fig (\ref{lolldual}). If we had inserted the operator at $y$ at the vertex 1 of the lollipop graph, we would have had an inequivalent situation. We need to consider this case also, since this graph does contribute to the spacetime correlator as well. The worldsheet correlator corresponding to this graph can be easily obtained by performing the required permutation in (\ref{gencorr}). This in general is a different function. On the other hand permuting vertices 2 and 3 as has been pointed out  earlier 
results in the same graph, so we could readily include the contribution by including an appropriate symmetry factor in the constant (\ref{constant}). However, we do have a further non-trivial consistency check of crossing symmetry. We will
address this issue  in the next section.

To conclude, we see that at large $J$ the world sheet correlator gets its only non-trivial contribution from the Jacobian $s \rightarrow \eta$, the map being given by the solutions of eqns (\ref{set1}) and (\ref{set2}) for fixed saddle point values of $s_2$ and $s_3$. We emphasize here that this result is
obtained only after we sum over the contributions from the whole family of lollipop diagrams. Since the contribution comes only from the Jacobian, we see that in this limit, the worldsheet correlator is somewhat universal. For instance apart from symmetry factors we would have obtained the same answer from a different integrand such as that of a matrix model. 

\section{ Worldsheet correlator for the exactly solvable point}

Implicitly, the world sheet correlator for the lollipop diagram
 is given by (\ref{gencorr}) in the large
$J$ limit.  To obtain the correlator explicitly we need to choose
 $x^2,y^2, z^2$ and $J_1, J_2, J_3$ such that the zeroth order equations in (\ref{pertsoln}) are solvable. However the saddle line depends only on two independent combinations of these variables, as the saddle line is fixed by saddle point values of $s_2$ and $s_3$ which are $\beta a/\alpha b$ and $\gamma a/\alpha c$ respectively. In other words the saddle line depends on the ratios of distances times the ratios of 
 the $R$-charges of the operators such that
there are just two independent parameters. 

We have seen that the equations (\ref{set1}) can be explictly solved for $s_2=s_3=1$.
In the large $J$ limit, it can be seen these values of $s_2, s_3$ are obtained by 
setting
$$x^2/ J_1 = y^2/ J_2 = z^2/J_3.$$ The cross ratio as a function of $s$ is 
given by (\ref{set2}):
\begin{equation}\label{saddleline1}
\eta = \eta_{o}(s) = -\omega \frac{ 1-\frac{B^2}{A^2}}{1-\omega^{2}\frac{B^2}{A^2}}, 
\qquad
\frac{B^2}{A^2}=\omega^{-\frac{1}{2}}\left(\tan(\frac{s}{4})\right)^{\frac{2}{3}} .
\end{equation}

To evaluate the world sheet correlator we need to determine
 $\bar{\eta_{o}}(\eta_{o}^{-1}(\eta)))$ and $\eta_{o}^{'}(\eta_{o}^{-1}(\eta))$. To do this we would not need to actually evaluate the function $\eta_{o}^{-1}$. We collect the following simple facts:
\begin{equation}\label{fact1}
\frac{B^2}{A^2} = \frac{\eta + \omega}{\eta\omega^2 + \omega} , 
\end{equation}
and
\begin{equation}\label{fact2}
\frac{\bar{B}^2}{\bar{A}^2} = \frac{1}{\omega}\frac{B^2}{A^2}.  
\end{equation}
The above equation follows from the second explicit expression in (\ref{saddleline1}). Combining (\ref{fact1}) and (\ref{fact2}) we obtain:
\begin{equation}
\frac{\bar{\eta}+\omega^2}{\bar{\eta}\omega+\omega^{2}}=\frac{\eta + \omega}{\eta  +\omega^{2}} .
\end{equation}
Solving the above we get $\eta\bar{\eta} = 1$. Thus the  ``saddle line'' is just the unit circle,  furthermore: 
\begin{equation}
\bar{\eta_{o}}(\eta_{o}^{-1}(\eta))) =\frac{1}{\eta} .
\end{equation}
Again using the second explicit expression in (\ref{saddleline1}) we can calculate:
\begin{eqnarray}\label{fact3}
\frac{\partial}{\partial s}\left( \frac{B^{2}}{A^{2}}\right) &=&
 \left(\frac{B^{2}}{A^{2}}\right)^{-\frac{1}{2}}\left(\frac{\omega^{\frac{-3}{4}}}{6}\right)
 \left[1 -\left(\frac{B^{2}}{A^{2}}\right)^{3}\right], \\\nonumber 
\frac{\partial \eta_{o}(s)}{\partial s} &=& \frac{\omega-1}{\left(1-\omega^{2}\frac{B^2}{A^2}\right)^{2}}\frac{\partial}{\partial s}\left(\frac{B^2}{A^2}\right).
\end{eqnarray}
Finally using  (\ref{saddleline1}), (\ref{fact1}) and (\ref{fact3}) we  obtain :
\begin{equation}
\eta_{o}^{'}(\eta_{o}^{-1}(\eta)) = \frac{1}{6}\frac{\omega^{-\frac{3}{4}}(\omega-1)}{\left(1-\omega^{2}\frac{\eta + \omega}{\eta\omega^{2} + \omega}\right)^{2}}\left(\frac{\eta + \omega}{\eta\omega^{2} + \omega}\right)^{-\frac{1}{2}}\left[1-\left(\frac{\eta + \omega}{\eta\omega^{2} + \omega}\right)^{3}\right] .
\end{equation}
Therefore the final explicit expression for our worldsheet correlator is:
\begin{equation}\label{explicitcorr}
\mathcal{G}(\eta,\bar{\eta}) \sim \delta(\bar{\eta} -\frac{1}{\eta})6(\omega -1)\omega^{\frac{3}{4}}(\eta\omega^{2}+\omega)^{-\frac{5}{2}}(\eta + \omega)^{\frac{1}{2}} 
\frac{1}{1-\left(\frac{\eta + \omega}{\eta\omega^{2} + \omega}\right)^{3}} .
\end{equation}
We observe that the worldsheet correlator in this case is a rational function of $\eta$. 
The constant of proportionality in (\ref{explicitcorr}) could be obtained as a special case of (\ref{constant})
\begin{equation}\label{constant2}
e^{-J\left[ \alpha \ln(\alpha) +\beta \ln(\beta) + \gamma \ln(\gamma)\right]  }
\left(\frac{2\pi} {J\sqrt{\alpha\beta\gamma} }  \right).
\end{equation}
This constant will be important when we add the contributions of the Y correlator. The  other lollipop contributions from the inequivalent graphs
are  obtained from the permutations of the vertices. In this limit,
the constant remains the same. It seems for this special case
 we have an accidental permutation symmetry among
$\alpha, \beta, \gamma$.

\subsection{Crossing Symmetry}

Now we will check if the world sheet  correlator 
in (\ref{explicitcorr}) exhibits crossing symmetry. 
This is indeed a non-trivial check of the proposal for construction of worldsheet correlators.  This check was performed
 for the Y diagram in \cite{GopakumarDavid}. 
 We will do the same here for the lollipop
 and show that the world sheet
 correlator in (\ref{explicitcorr}) 
 exhibits crossing symmetry.
  On drawing  the lollipop on the sphere it is clear the graph has the property that the edge connecting the second and fourth vertices and that connecting the third and the fourth vertices are equivalent. 
 Therefore this equivalence should be reflected in the worldsheet correlator also.

The world sheet correlator in the
large $J$ limit  is localised on a line in the complex plane and so we should look
only for those $SL(2,C)$ transformations which keep the line invariant. 
Let $f(z)$ be 
the  conformal tranformation
which maps the saddle line to the real line.
The stabilizer of the real line is $SL(2,R)$. Then the
desired subgroup of $SL(2,C)$ which keeps our saddle line invariant is
the conjugation $f^{-1}\cdot SL(2,R)\cdot f$. 

To verify crossing symmetry we need to find the $SL(2,C)$ transformation that interchanges $v_{2}$ and $v_{3}$ and then check if this is part of the subgroup of $SL(2,C)$ which stabilizes our saddle line. The transformation is $\eta \rightarrow 1/\eta$. Indeed, since the saddle line is the unit circle, this transformation does stabilise it.

More explicitly, let us examine the effects of  the 
transformation 
$\eta \leftarrow 1/\eta$ on two parts of  (\ref{explicitcorr}) individually.
 Firstly let us look at the delta function part. It transforms to the expression below:
\begin{equation}\label{cross1}
\delta(\frac{1}{\bar{\eta}}-\eta) =-\bar{\eta}^{2}\delta(\bar{\eta}-\frac{1}{\eta}).
\end{equation}
Here we have rewritten the delta function by thinking of it as a ``primary''
function of $\bar\eta$, i.e. we are going to integrate it out 
first.
The remaining part of the correlator (\ref{explicitcorr}) is given by:
\begin{equation}\label{cross2}
f(\eta)=6(\omega -1)\omega^{\frac{3}{4}}(\eta\omega^{2}+\omega)^{-\frac{5}{2}}(\eta + \omega)^{\frac{1}{2}} 
\frac{1}{1-\left(\frac{\eta + \omega}{\eta\omega^{2} + \omega}\right)^{3}} .
\end{equation}
It is straightforward to verify that:
\begin{equation}
f(\frac{1}{\eta})=-\eta^{2}f(\eta).
\end{equation}
To complete the check we need to see 
Finally 
it is easy to see that the multiplicative constant  in (\ref{constant2}) 
is also invariant under simultaneous exchange of $b,c$  and $J_{2}, J_{3}$.
Therefore, from (\ref{cross1}) and (\ref{cross2}) we obtain
\begin{equation}
\mathcal{G}_{y,J^2 ; z,J^3}(\frac{1}{\eta},\frac{1}{\bar{\eta}}) = \vert\eta\vert^{4}\mathcal{G}_{z,J^3 ; y,J^2}(\eta,\bar{\eta}).
\end{equation}
Thus the world sheet correlator indeed has crossing symmetry.

Generically, we also have other inequivalent contributions, for instance when we exchange vertices 1 and 2. These would not be related to each other by crossing symmetry. These inequivalent contributions would have support from different saddle lines because they would not be stabilised under the $SL(2,C)$ transformations which would implement the permutations. Hence the whole worldsheet correlator gets support from these 
complete set of saddle lines obtained from all the inequivalent lollipop graphs.

\section{Discussion}

We have observed in section 3,  that in the limit of large $J$ (with ratios fixed), the planar worldsheet four point correlator coming from the lollipop graphs, gets its non-trivial contribution only from the Jacobian of the change of variables from the Schwinger parameters to the worldsheet moduli space. The integrand contributes just through its value at the saddle point of the set of variables on which it depends. This happens, however, very non-trivially, only when we sum over a whole family of diagrams. Since in this limit the integrand doesn't contribute, there is a measure of universality to the answer. For instance the worldsheet  correlator would have been the same had the integrand  been that of a matrix model. It would be interesting to understand the significance of this a bit better.

Since by the operator state correspondence in AdS/CFT, the large $J_i$ limit is dual to highly massive Kaluza-Klein states in the dual string theory. We have also taken the large $N$ limit so that we need consider only classical string theory. If it is indeed the case that the world sheet correlators of these extremal four point functions are not renormalized with 't Hooft coupling, then we could try and compare these correlators with answers at large 't Hooft coupling.  It would be nice to see if a geometric approach to the scattering of these states in the large radius $AdS$ could show the signature of the above localisation onto a curve in the moduli space.

\acknowledgments We would like to thank  ICTS, TIFR for hospitality while completing this manuscript. One of us (A. M.) would also like to thank IACS, Kolkata for hospitality. Finally, we are grateful to the people of India for generously supporting  research in string theory.

 \bibliography{lolli}
\bibliographystyle{JHEP}

\end{document}